\documentclass[10pt,journal, compsoc]{IEEEtran}
\usepackage{amsmath,amsfonts}
\usepackage[caption=false,font=normalsize,labelfont=sf,textfont=sf]{subfig}
\usepackage{textcomp}
\usepackage{stfloats}
\usepackage{url}
\usepackage{verbatim}
\usepackage{graphicx}
\usepackage{cite}
 \usepackage{times}                     % we use Times as the main font
\usepackage{tabu}                      % only used for the table example
\usepackage{booktabs}                  % only used for the table example
\usepackage{lipsum}                    % used to generate placeholder text
\usepackage{mwe}                       % used to generate placeholder figures
\usepackage{multirow}
\usepackage{mathptmx}                  % use matching math font
\usepackage{glossaries}
\usepackage{graphicx}
\usepackage{xspace}                    % used to manage spacing at the end of macros (commands.tex)
\usepackage{svg}
\usepackage{verbatim}

\usepackage{placeins}
\usepackage{adjustbox}
\usepackage{mathptmx}                  % use matching math font
\usepackage{glossaries}
\usepackage{graphicx}
\usepackage{xspace}                    % used to manage spacing at the end of macros (commands.tex)
\usepackage{svg}
\usepackage{verbatim}

\usepackage{placeins}
\usepackage{adjustbox}

  \usepackage{enumitem}
 \usepackage{rotating}
\usepackage{longtable}
\usepackage{acro}
\usepackage{algorithm}
\usepackage{algorithmicx}
\usepackage[algo2e]{algorithm2e}
\usepackage{algcompatible}
\usepackage{wrapfig}
\usepackage{textcomp}
\usepackage{xspace}
\usepackage{amsmath}
\usepackage{setspace}
\let\Algorithm\algorithm
\renewcommand\algorithm[1][]{\Algorithm[#1]\setstretch{1.5}}
\usepackage{placeins} % limiting where the floats can be.
\usepackage{enumitem}
\usepackage{balance}
\usepackage{lineno}

\usepackage{subfig}
\usepackage{multirow} 

\usepackage{graphicx}
\usepackage{mathptmx}                  % use matching math font
\usepackage{amsmath}%new 
\usepackage{amsfonts}

\usepackage{xcolor}
\usepackage[normalem]{ulem}

\usepackage{adjustbox}

\makeglossaries

\newacronym{llm}{LLM}{Large Language Model} 
\newacronym{asr}{ASR}{Automatic Speech Recognition}
\newacronym{adb}{ADB}{Android Debug Bridge}
\newacronym{gguf}{GGUF}{GPT-Generated Unified Format}
\newacronym{ui}{UI}{User Interface}
\newacronym{nn}{NN}{Neural Network}
\newacronym{pp}{PP}{Prompt Processing}
\newacronym{tg}{TG}{Token Generation}
\newacronym{bt}{BT}{Batch Test}
\newacronym{tt}{TT}{Thread Test}
\newacronym{rss}{RSS}{Resident Set Size} 
\newacronym{avp}{AVP}{Apple Vision Pro} 
\newacronym{mq3}{MQ3}{Meta Quest 3} 
\newacronym{ml2}{ML2}{Magic Leap 2} 
\newacronym{vivo}{Vivo}{Vivo X100 Pro}

\begin{document}
\title{GHAR: GeoPose-based Handheld Augmented Reality for Architectural Positioning, Manipulation and Visual Exploration}
%\title{LLMs on XR (LoXR): Performance Evaluation of LLMs Executed Locally on XR Devices }
  \vskip -0.365cm
\author{ Sabahat~Israr, Dawar~Khan$^*$, Zhanglin Cheng, Mukhtaj Khan, Kiyoshi Kiyokawa

\IEEEcompsocitemizethanks{ 
\IEEEcompsocthanksitem Sabahat~Israr and Mukhtaj Khan are with The University of Haripur, Haripur, Pakistan. 
\IEEEcompsocthanksitem Dawar~Khan is with King Abdullah University of Science and Technology (KAUST), Saudi Arabia. E-mail: dawar.khan@kaust.edu.sa. 
\IEEEcompsocthanksitem Zhanglin Cheng is with Shenzhen Institute of Advanced Technology, Chinese Academy of Sciences, Shenzhen China.
\IEEEcompsocthanksitem Kiyoshi Kiyokawa is with Nara Institute of Science and Technology, Nara, Japan.
 }
\thanks{Manuscript received MM dd, YYYY; revised MM dd, YYYY.\\ ($^*$ Corresponding author: Dawar Khan.)}}

        % <-this % stops a space

\markboth{For arXiv preprint, June~2025}%
{Shell \MakeLowercase{\textit{et al.}}: A Sample Article Using IEEEtran.cls for IEEE Journals}

\IEEEpubid{0000--0000/00\$00.00~\copyright~2021 IEEE}

\IEEEtitleabstractindextext{ 
\begin{abstract} 
Handheld Augmented Reality (HAR) is revolutionizing the civil infrastructure application domain. The current trend in HAR relies on marker tracking technology. However, marker-based systems have several limitations, such as difficulty in use and installation, sensitivity to light, and marker design. In this paper, we propose a markerless HAR framework with GeoPose-based tracking. We use different gestures for manipulation and achieve 7 DOF (3 DOF each for translation and rotation, and 1 DOF for scaling). The proposed framework, called GHAR, is implemented for architectural building models. It augments virtual CAD models of buildings on the ground, enabling users to manipulate and visualize an architectural model before actual construction. The system offers a quick view of the building infrastructure, playing a vital role in requirement analysis and planning in construction technology. We evaluated the usability, manipulability, and comprehensibility of the proposed system using a standard user study with the System Usability Scale (SUS) and Handheld Augmented Reality User Study (HARUS). We compared our GeoPose-based markerless HAR framework with a marker-based HAR framework, finding significant improvement in the aforementioned three parameters with the markerless framework.
\end{abstract}
\begin{IEEEkeywords}
Markerless Handheld Augmented Reality, Large Scale Applications, GeoPose, 3D manipulation, 3D Visualization, Architecture
\end{IEEEkeywords}
 } 
\maketitle

% \begin{abstract}
% This document describes the most common article elements and how to use the IEEEtran class with \LaTeX \ to produce files that are suitable for submission to the IEEE.  IEEEtran can produce conference, journal, and technical note (correspondence) papers with a suitable choice of class options. 
% \end{abstract}

% \begin{IEEEkeywords}
% Article submission, IEEE, IEEEtran, journal, \LaTeX, paper, template, typesetting.
% \end{IEEEkeywords}

%\section{Introduction}
%\IEEEPARstart{T}{his} 

 \IEEEdisplaynontitleabstractindextext 
 \IEEEpeerreviewmaketitle
%\maketitle
% % Use this for peer review papers
% \IEEEpeerreviewmaketitle
\IEEEraisesectionheading{\section{Introduction}}
\label{intro} 
Manufacturing and construction companies are investing billions of dollars in the industrial metaverse. In this regard, handheld devices are available as the relatively cheaper solutions.  These devices provide a range of applications, making a positive impact on both large-scale industrial activities and smaller-scale ventures. In scientific research, particularly in fields like training, medicine, civil engineering, and education, handheld devices play a crucial role. They provide real-time access to information, ensuring ease of handling and use. To unlock their full potential, various visualization techniques, including three-dimensional display methods, are employed to present knowledge or projects effectively. Augmented reality (AR) stands out as one of the widely adopted visualization techniques in this context\cite{BIMAR}.\\
%\begin{figure*}[t]
%\centering
%\centerline{\includegraphics[height=1.4in,width=5.5in]{images/cont.png}}
%\caption{Reality virtuality continuum:  From left to right: Real environment (all real contents), AR (virtual objects overlaying the RE), Augmented Virtuality (Real object inside a VE), and VR (pure computer generated)~\cite{RVConsortiuum,ROMPAPAS201924,URLASTAR}.}
%\label{fig:VRcont}
%\end{figure*} 
AR is the technological advancement that combines both real and virtual world by augmenting computer generated perceptual content to the real world environment, thus providing real-time 3D interaction \cite{azuma}. It lies near the real world environment on the reality virtuality continuum. AR technology   %(see~\autoref{fig:VRcont})~\cite{RVConsortiuum}
has many use cases in different areas including medical \cite{medical}, tourism \cite{tourism}, manufacturing \cite{manufacturing}, marketing \cite{marketing}, and architecture \cite{architecture}, etc.  AR has undergone rapid development in civil infrastructure \cite{civil1, civil2, civil3} and still holds potential for breakthroughs in unexplored areas within this domain. Researchers commonly apply AR in civil infrastructure in two primary ways: firstly, in the design phase, encompassing structures like bridges, railways, and buildings; and secondly, in various application scenarios such as architecture design, structural design, construction, and more \cite{Arc1, Arc2, Arc3}.\\
AR is categorized into marker-based tracking and markerless tracking \cite{TypesofAR}. Marker-based tracking uses artificial markers (fiducials) as trackables, while markerless tracking relies on real environmental features \cite{markerlessFeatures} for overlaying virtual content onto the real scene. Despite being accurate and easy to implement, marker-based tracking has limitations, including susceptibility to light conditions, the need for proper marker design, and the requirement for markers to be distinct from each other \cite{ARToolkit, RobustTracking}.\\
After tracking, users set an anchor point, determining a location for 3D virtual content. Users are granted degree of freedom (DOF) to manipulate position, orientation, and scaling. DOF cover translation and orientation of real or virtual objects, with translation along x, y, and z axes, and orientation around x, y, and z axes \cite{DOF}. Moreover, uniform scaling corresponds to 1 DOF \cite{Poster}. In AR applications, users employ various gestures or techniques to manipulate degree of freedom (DOF) effectively. \\
 AR helps civil engineers and construction workers in planning and surveying, offering efficient visuals and interactive modifications. In congested or restricted areas, new construction impacts traffic and desired structures. AR visualizations help in visualizing changes and act as a bridge of communication among the owners, engineers, and construction teams (including the laborers). It minimizes uncertainties and addresses communication gaps among the stakeholders. The significance of AR in civil engineering is clear, yet convincing the community to adopt it remains a challenge. To address this, an easy-to-use AR system is crucial, prioritizing user satisfaction and ease of manipulation. Incorporating different degrees of freedom, and a simple interface enhances user satisfaction and ease of manipulation. In summary, end users, particularly in the context of civil engineering, encounter challenges in both tracking and post-tracking phases of Handheld AR.\\
 This paper introduces a framework named \textit{GHAR}\footnote{\textit{GHAR}~(\raisebox{-1.975pt}{\includegraphics[height=8pt]{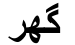}}) means “home” in the Urdu language.} (GeoPose-based Handheld Augmented Reality).
 GHAR enables marker-less handheld augmented reality through GeoPose \cite{Kresse2022}, facilitating the positioning and placement of virtual objects, specifically architectural models of buildings, with respect to the globe. Geospatial API is used to implement Geopose anchors. It provides high accurate positioning and real-time localization in challenging visual environments as it uses visual cues and reference data from 3D models or satellite imagery. GeoPose leverages advanced computer vision algorithms to estimate the camera pose based on a combination of image features, 3D models, and known landmarks which helps in handling dynamic scenes, changes in lighting conditions, and varying camera viewpoints robustly as compared to Global Positioning System (GPS). A brief comparison between geospatial anchors and GPS location anchor is described in Table \ref{tab:comparison}. 
% Table generated by Excel2LaTeX from sheet 'Sheet3'
\begin{table}[htbp]
  \centering
  \caption{Comparison between Geospatial Anchor and GPS Location Anchor.}   
\resizebox{\linewidth}{!}{
\begin{tabular}{p{6em}p{9.5em}p{9.5em}}
    \hline
    \textbf{Feature} & \textbf{Geospatial Anchor} & \textbf{GPS Location Anchor} \\
    \hline
    \textbf{Technology} & ARCore Geospatial API and VPS (Visual Positioning System) & Relies purely on GPS and standard location services \\
    \hline
    \textbf{Accuracy} & High (within a few centimeters) & Low to moderate (typically 3-10 meters) \\
    \hline
    \textbf{Environmental Context} & Leverages visual data from surroundings (e.g., buildings, signs) & Based on satellite signals, independent of surroundings \\
    \hline
    \textbf{Tracking} & Requires camera tracking and AR session & GPS tracking only (no camera needed) \\
    \hline
    \textbf{3D Spatial Anchoring} & Anchors in 3D space (latitude, longitude, altitude, heading) & Anchors primarily in 2D (latitude, longitude) \\
    \hline
    \textbf{Persistence} & Can persist in 3D space and reappear when revisited & No persistence of orientation, limited altitude accuracy \\
    \hline
    \textbf{Reliability in Urban Areas} & High (better with landmarks and visual cues) & Less reliable due to GPS signal interference \\
    \hline
    \textbf{Orientation and Heading} & Provides accurate heading and orientation & Low accuracy in Heading; needs compass. \\
    \hline
    \textbf{Dependency on VPS} & Yes – Visual positioning is required & No \\
    \hline
    \end{tabular}%
  \label{tab:comparison}%
  }
\end{table}%

To the best of our knowledge this is the first 7DOF HAR framework that uses GeoPose based marker-less tracking. Compared with existing techniques, we have three main advantages: (1) markerless tracking, (2) 7-DOF interaction, and (3) implementation on large-scale application using handheld devices. Moreover, it helps in real world problem i.e., the utilization of the proposed framework for architectural and civil engineering.\\ 
The key contributions are enlisted below. 
\begin{itemize}[noitemsep]
\item Proposed a marker-less Handheld Augmented Reality (HAR) framework which uses GeoPose for tracking and pose estimation. The framework retains realism and tracking accuracy with consideration of usability, comprehensibility and manipulability. 
\item Achieved 7-DOF manipulation of 3D objects (i.e. rotation (3), translation (3), and scaling (1)) with minimal user efforts which makes our system suitable for large scale augmented reality application. 
\item Application of the proposed framework for modelling and automation in the architectural and civil engineering designs.
\item  Performed a comprehensive user study and evaluated the usability, manipulability, and comprehensibility of the proposed system using standard methods i.e., HARUS~\cite{HARUS} and SUS~\cite{SUS}.  
\end{itemize}

 The remaining sections of the paper are structured as follows: Section \ref{intro} provides an introduction to the research topic, elucidating its significance and the rationale behind its pursuit. Section \ref{sec:rw} delves into the existing literature, while Section \ref{proposedFramework} outlines the methodology of our proposed framework. Section \ref{evaluation} details the evaluation paradigm, and Section \ref{results} elucidates the experimentation results and their analysis. In Section \ref{discussion}, the findings of the results are discussed and reasoned. Finally, Section \ref{conclusion} concludes our research work, offering insights into potential future directions.

\section{Related Work}
\label{sec:rw} 
 In HAR, 3D interaction with digital assets comprises different manipulation tasks such as selection, translation, orientation, and scaling \cite{userInterfaces, Zoom-fwd}. Efficient and accurate 3D object manipulation serves as the backbone for the spatially consistent alignment of real and virtual objects, a key element of augmented reality applications \cite{azumaManipulate}. This section thoroughly explores various interaction techniques and methodologies used for 3D object manipulation in HAR. Researchers aim to make interactions easy for users in the augmented reality environment, categorizing interaction techniques into touch-based interaction \cite{interaction1, interaction2}, mid-air gestures-based interaction \cite{midAir1, midAir2}, and device-based interaction \cite{device1}.

Numerous methodologies have been proposed to make interactions user-friendly in the augmented reality environment. SlidAR \cite{Polvi} is a manipulation technique where researchers used ray casting and epipolar geometry for the accurate positioning of virtual content in the real world. This technique employed Simultaneous Localization and Mapping (SLAM) for positioning. SlidAR+ \cite{SlidAR+} is an extension of SlidAR \cite{Polvi}, utilizing its SLAM-based positioning method. The main idea was to pre-align virtual content either parallel or perpendicular to the gravity vector during initialization. For orientation tasks, one rotation axis would always be aligned with the direction of gravity for fast adjustments.

The DS3 technique \cite{DS3} separated the control of orientation and translation. It used the Z-technique proposed by \cite{DS3Translate} to control the position of the object and the constraint solver proposed by \cite{screenSpace} to control orientation. Translation was performed using one direct finger, except for depth translation, which was performed indirectly using a second finger. Orientation was controlled with at least two direct fingers in an integral way using a constraint solver. Screen Space \cite{screenSpace} was used for the direct manipulation of 3D objects with three or more points, ensuring a controllable mapping between points in screen-space and points in the local space of an object.

Similarly, Shallow-Depth \cite{ShallowDepth} offers 3D interaction in the z-plane. The authors also proposed guidelines for direct touch 3D interaction and demonstrated their significance with a user study. Another technique proposed in \cite{6DOF} manipulated the 6DOF of 3D objects with only two-finger gestures, interpreting the moving characteristics of fingers to manipulate panning, pinching, swiveling, and pin-panning. The performance was compared with other state-of-the-art techniques \cite{screenSpace, DS3}, showing it to be least sensitive to screen and object sizes and outperforming other techniques in terms of task completion time.

Moreover, \cite{multiTouch} proposed a multi-touch-based 3D interaction technique, allowing users to manipulate virtual objects using a combination of different gestures resulting in three different interactions: translation control, orientation control, and scale control. Gestures that could be combined for manipulation include tap, press, and drag. The Zoom-IN technique \cite{ZoomIn} used the camera zoom feature to select an object, aiming to bring two distant virtual and real objects close together while maintaining spatial registration. A virtual hand metaphor was used for object manipulation. Similarly, \cite{Poster} presented three methods for 3D object manipulation: direct manipulation via touch gestures, indirect manipulation via uni-manual and bi-manual thumb touch gestures. This technique was implemented on a large-sized mobile device rather than smaller screen smartphones, tablets, or handheld devices.\\
%\section{Problem Statement}
%\label{ps}
% Table generated by Excel2LaTeX from sheet 'Sheet2'
\begin{table*}[!htbp]
%\vskip -0.5cm
  \centering
  \caption{Description of state-of-the-art manipulation techniques and their limitations.}
  %\begin{adjustwidth}{-\extralength}{-3cm}
 %   \begin{tabular}{|c|p{4.235em}|p{9.265em}|r|r|r|r|p{9.235em}|r|r|r|r|}
  % \resizebox{\linewidth}{!}{%
  \begin{tabular}{p{13em}p{20em}p{18em}}
    \hline
%     \multicolumn{1}{|l|}{\textbf{NO.}} & \multicolumn{2}{p{8.1em}|}{\textbf{3D Manipulation Techniques}} & \multicolumn{2}{c|}{\textbf{Description}} & \multicolumn{4}{\textbf{Limitations}}  \\
     \textbf{Technique} & {\textbf{Description}} & \textbf{Limitations}  \\
    \hline
   SlidAR+ \cite{SlidAR+} & Used raycast and epipolar geometery for positioning 3D content and gravity vector constraint for orientation control & Confined to marker, not applied to real world application, didn't take into account physical limitations of device \\
   \hline
    SlidAR \cite{Polvi} & Used raycast and epipolar geometery for positioning 3D content & Confined to marker, not applied to real world application, didn't take into account physical limitations of device \\
    \hline
    Interactions with 3D virtual objects in AR using natural gestures \cite{InteractAR} & Used two-stage palm detection to overlay 3D virtual content, one-hand (uni-manual) gestures are used to interact with augmented content & Had computational complexity due to complex algorithms used, had recognition and rendering problem i.e system couldn't recognize few hand movements \\
    \hline
    6DOF Manipulation by Two-Fingers \cite{6DOF}& Two-finger gestures for 6-DOF control on small display & Used single touch device only \\
    \hline
    Depth-Separated Screen-Space \cite{DS3} & Provided separated control of translation and rotation  & Imposed more cognitive load  \\
    \hline
    Sky-line based Approach \cite{skyline}& Used skyline pixels to estimate the pose and skyline acted as a marker to overlay content & Had physical limitations and computational complexity due to complex algorithms  \\  
    \hline
    Position-based AR for aiding construction and inspection of offshore plants \cite{PositionOffshorePlants}& Used accelerometer and gyro for pose estimation, aligned actual object with 3D CAD model and augment information & Non-robust tracking based on visual features, Used extra sensors for pose estimation and tracking, confined to marker\\
    \hline
    Spatial augmented reality based circuit experiment \cite{SpatialAR}& Virtual experiment phenomenon of circuit experiment like brightness change of light bulb and flow of charge in a circuit, was visualized and displayed in physical word & This technique couldn't be used in outdoor applications due to instabilities and its requirement for projector, stable illumination and proper hardware setup \\
    \hline 
    Shallow-Depth \cite{ShallowDepth}& Uses 3 fingers gestures to control all 6 DOFs &Complex, Impose cognitive load\\
    \hline
    Scree Space \cite{screenSpace} & Direct manipulation of 6DOFs with 3 or more fingers & No results about user experiences.\\
    \hline
    3DTouch \cite{3DTouch} & Uses one and two finger gestures to scale, rotate and translate 3D object as integral operations. & Has physical limitations \\
    \hline
    Multi-touch 3D interaction \cite{multiTouch} & Uses one and two finger gestures for translation, orientation and scale control& Not tested for any user experience \\
    \hline
    Zoom-IN \cite{ZoomIn} & Uses camera zoom feature to select an object and virtual hand metaphor to manipulate it & Is costly, implemented on head mounted displays  \\
    \hline
    Poster \cite{Poster} &Three techniques for direct/indirect 3D objects manipulation via thumb gestures & Implemented on large sized mobile device \\
    \hline
%    \multicolumn{1}{|l|}{\textbf{NO.}} & \multicolumn{2}{c|}{\textbf{Proposed Technique}} & \multicolumn{5}{c|}{\textbf{Description}} & \multicolumn{5}{c|}{\textbf{Limitations}} & \multicolumn{1}{p{8.201em}|}{\textbf{Implementation on large-scale application }} \\
    \hline
    GHAR (Proposed framework) & Uses GeoPose and fiducial markers to overlay and place 3D virtual content. Pinch, slide and twist gestures are used to scale, translate and rotate 3D virtual content. User can visually explore the 3D artifact by moving around it in 360    & Used by single user without any collaborative support. User study was conducted with participants having less familiarity with AR.  \\
    \hline
    \end{tabular}%}%
    %\end{adjustwidth}
  \label{tab:literature}%
%\vskip -0.5cm
% }
\end{table*}%  

 Researchers have proposed various techniques for 3D localization and manipulation in HAR with the primary aim of simplifying the placement and manipulation of 3D virtual objects for end users of HAR devices. However, these techniques exhibit limitations, such as a lack of robustness in dynamic environments \cite{3DTouch, skyline}, localization inconsistencies \cite{Polvi}, and inappropriateness for large-scale applications and environments \cite{SlidAR+}. Some approaches also lack comprehensive user testing \cite{screenSpace, multiTouch}, which is essential for proper evaluation. Many existing techniques in the literature focus on proposing gestures and methods to enhance manipulation without considering tracking accuracy and physical limitations, such as the computation power of the device \cite{Polvi, 3DTouch, Poster}.

A comparison between our proposed technique and previously suggested methods for manipulating 3D virtual content is presented in Table \ref{tab:literature}. To the best of our knowledge, the most relevant works are SlidAR+ \cite{SlidAR+} and SlidAR \cite{Polvi}. However, these methods are marker-based and limited to 6DOF. Moreover, there is no specific application use case in architectural design or civil engineering.

\section{Proposed Framework}
\label{proposedFramework} 

This study introduces a practical Handheld Augmented Reality (HAR) framework aimed at improving three core aspects: \textit{manipulability}, \textit{comprehensibility}, and \textit{usability}. We define two design goals:
\begin{itemize}[noitemsep]
    \item Simplify the HAR system so users can focus on their task rather than managing markers or tracking stability.
    \item Enable accurate and robust placement of 3D virtual content in diverse environments.
\end{itemize}
The three key factors targeted in our design are:
\begin{itemize}[noitemsep]
    \item \textbf{Manipulability:} Refers to how easily users can interact with and control virtual elements in the augmented environment.
    \item \textbf{Comprehensibility:} Relates to how clearly users understand and interpret the AR information presented.
    \item \textbf{Usability:} Denotes the overall ease of use and user satisfaction with the system.
\end{itemize}
To address these goals, we propose \textit{GHAR}—a markerless HAR framework enabling accurate placement and intuitive 7-DOF manipulation using GeoPose. GeoPose provides both geospatial position and orientation, ensuring stable and drift-free AR content anchoring, similar to GPS-level accuracy.

Users can manipulate AR models using one- and two-finger gestures (slide, pinch, twist) for translation, scaling, and rotation. The system supports immersive visualization and interaction with 3D architectural models placed in real-world scenes, aiding construction and design workflows.

\textit{GHAR} is implemented using Unity3D, Microsoft Visual Studio, ARFoundation, and architectural models sourced from SketchUp’s 3D Warehouse\footnote{\url{https://3dwarehouse.sketchup.com/?hl=en}}.

\begin{figure*}[t]
\centering
\includegraphics[width=\linewidth]{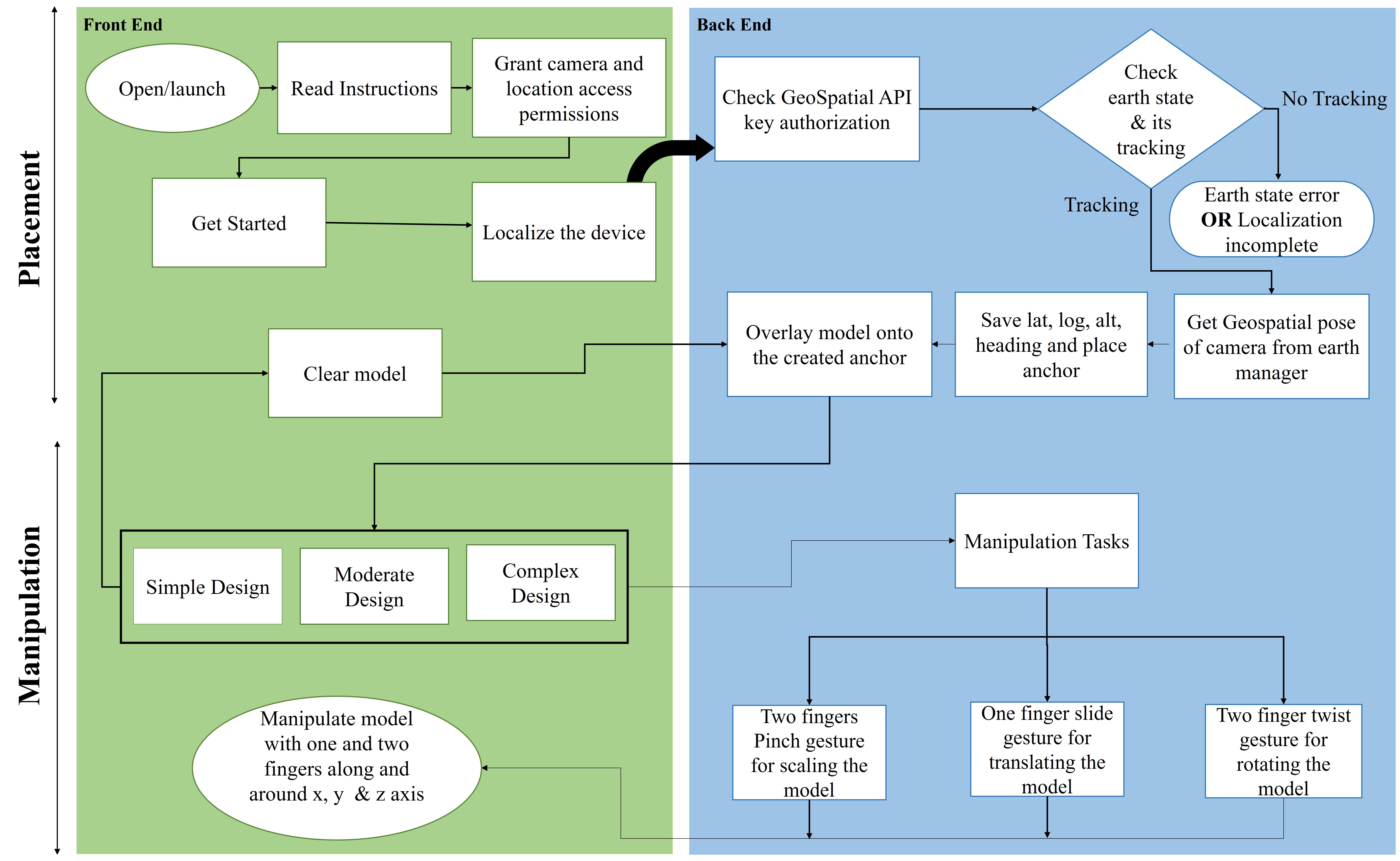}
% \centerline{\includegraphics[height=4.5in, width=6.9in]{images/Architecture.jpg}}
\caption{The overall pipeline of the proposed \textit{GHAR}.}
\label{fig: GHAR}
\end{figure*} 
\subsection{Positioning with respect to GeoPose}
We have used Geospatial API to place 3D models with respect to the globe. We used built-in sensors of the handheld device to determine the camera pose as shown in equation 1, where $P_C$ represents camera geospatial pose (latitude, longitude, altitude, Heading).
\begin{equation}
    P_c = (lat,lon,alt,\theta)  \label{eq1}
\end{equation}
After that we have computed the camera rotation based on its heading as shown in equation \ref{eq2} where Q represents adjusted camera rotation and $\theta_C$ represents heading/orientation of camera in degrees. 
\begin{equation}
    Q = Q(180^{\circ} - \theta_C)     \label{eq2}
\end{equation}
In the next step, anchor is placed at the computed pose as shown in equation \ref{eq3}.
\begin{equation}
    A = Anchor_{Add}(lat,lon,alt,Q)   \label{eq3}
\end{equation}
Then the building model is overlaid onto the placed anchor. The overall pipeline is illustrated in Figure \ref{fig: GHAR}. At the front-end, when a user launch/open the application, an instruction panel is prompted on the screen which guides the user about the use of the framework. After getting started and granting camera and location access permissions, user have to scan the environment around so that the camera can capture buildings, stores or any other visual elements around in order to localize the device as described in above equations. 
%Localization results in determining the latitude, longitude, altitude and heading values of where the device is located so that the user can place the model by setting an anchor.
\begin{figure*}[t]
\includegraphics[width=0.323\linewidth]{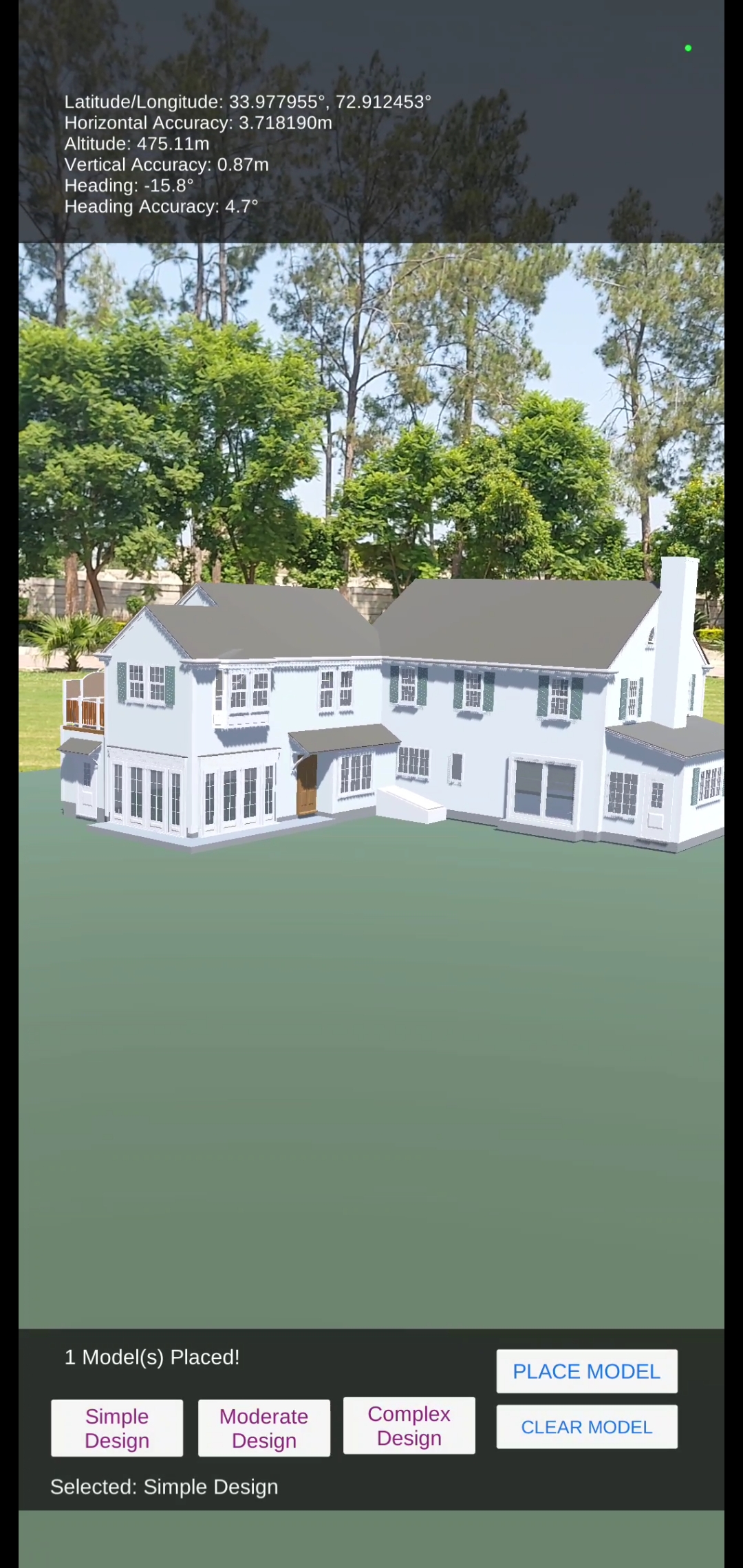}
\includegraphics[width=0.323\linewidth]{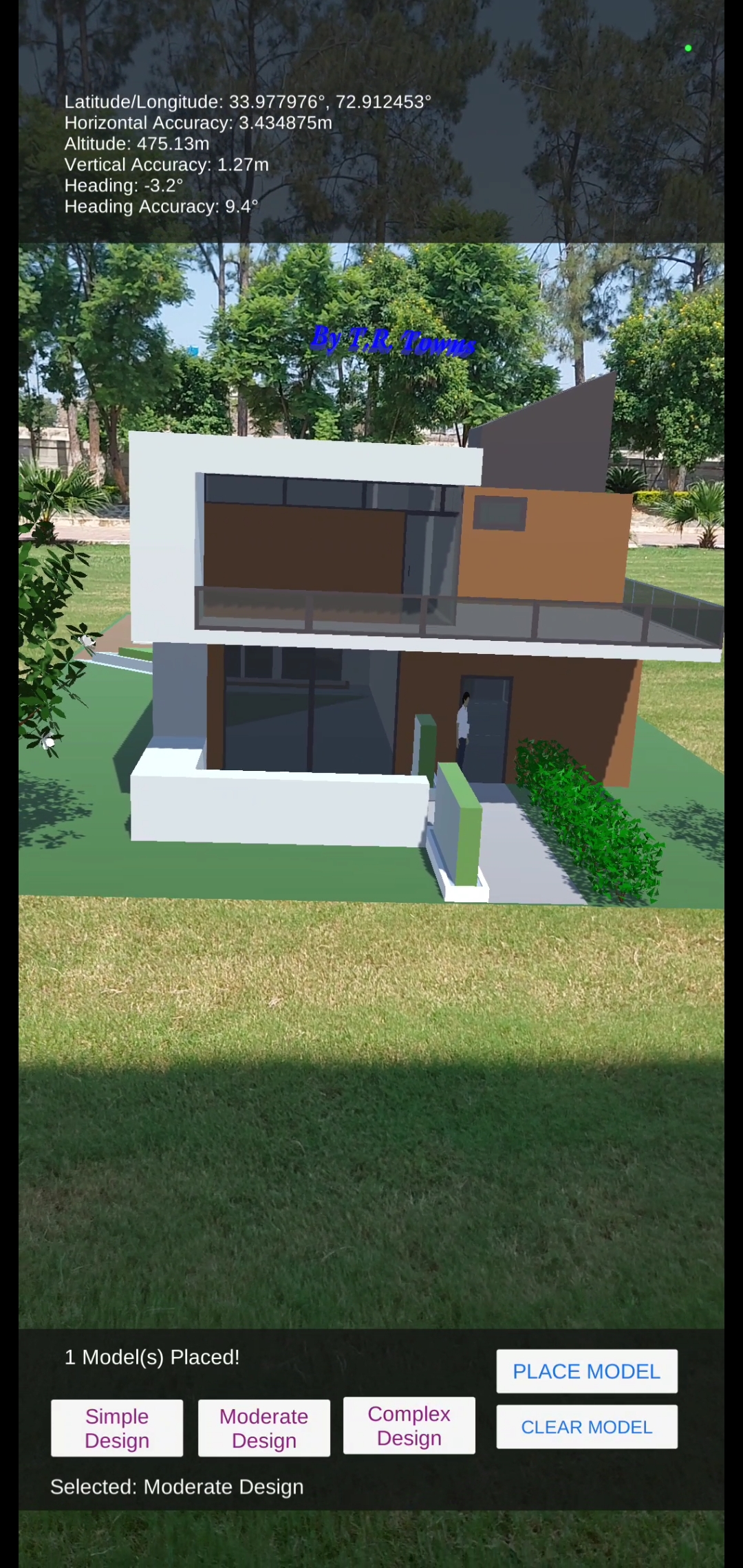}
\includegraphics[width=0.323\linewidth]{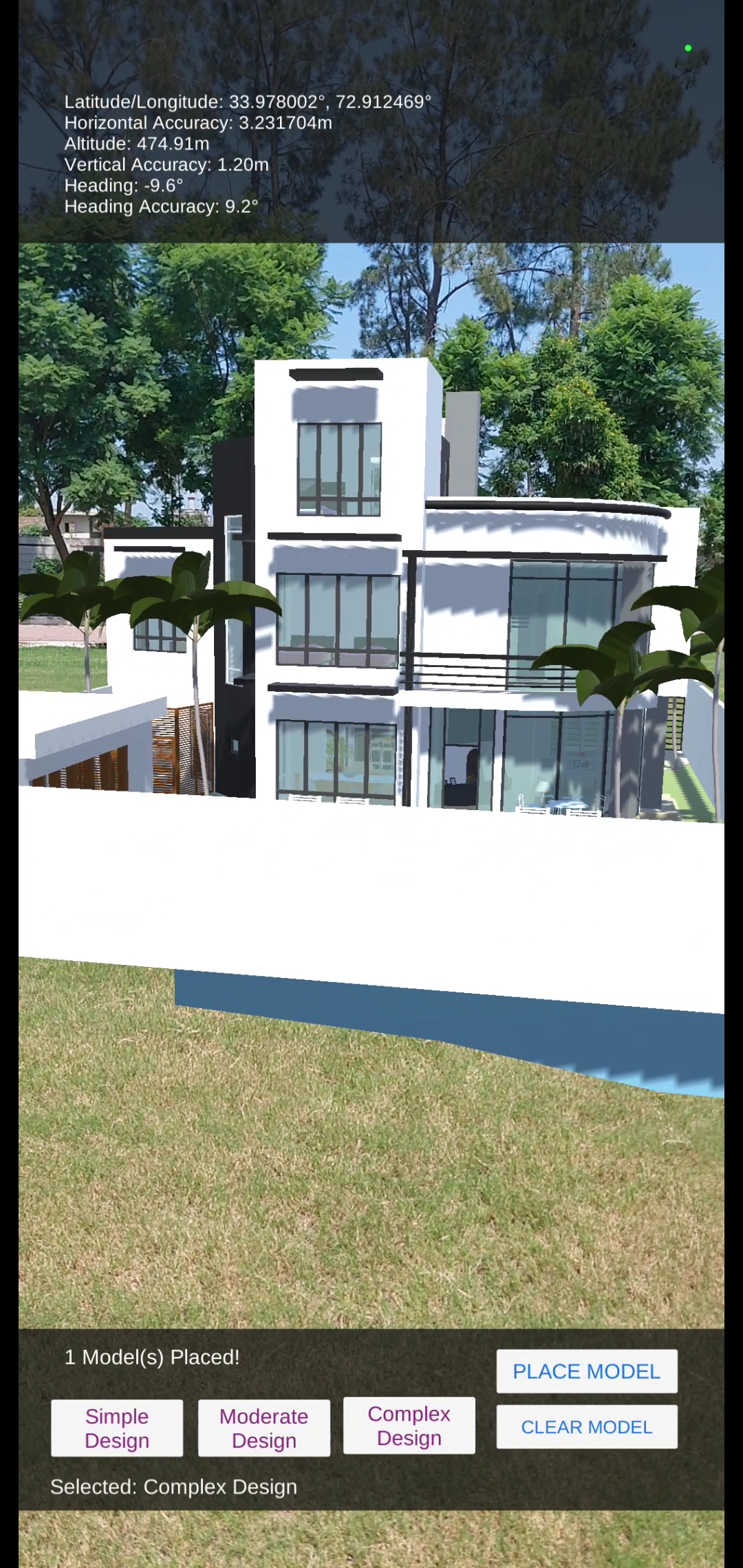}
% \begin{minipage}[c][4.8in]{2.3in}
% \vspace*{\fill} 
% \centering
% \includegraphics[height=4.8in, width=2.28in]{images/SD2.jpg}
% \end{minipage}%
% \begin{minipage}[c][4.8in]{2.3in}
% \vspace*{\fill}
% \centering
% \includegraphics[height=4.8in, width=2.28in]{images/MD2.jpg}
% \end{minipage}
% \begin{minipage}[c][4.8in]{2.3in}
% \vspace*{\fill}
% \centering
% \includegraphics[height=4.8in, width=2.29in]{images/CD2.jpg}
% \end{minipage}
\caption{3D building models placed on geographically anchored pose of device.}
\label{fig:GeoPosePlacement}
\end{figure*} 

\subsection{Working mechanism of GeoSpatial API}
The device establishes a connection to a visual positioning system by making a server call. Using the device's geographic coordinates and images, it compares them to high-resolution data from Google Maps. This process, known as visual localization, accurately determines the device's location in relation to its surroundings, surpassing the precision of GPS. The foundation of this system relies on Google's extensive collection of street view images captured worldwide over several years. Persistent visual features are described and identified for long-term recognition. This localization model encompasses an immense number of reference points, spanning across all countries with street views. When the device sends a request to the GeoSpatialAPI, a similar process is applied to the provided image. Pixels are analyzed, identifying recognizable points, and are matched with the localization model. The device's position and orientation is then calculated, which are subsequently relayed back to the user. GeoSpatialAPI combines the device's local coordinates with the corresponding geographical coordinates, enabling seamless work within a unified coordinate system and set an anchor at lat/long/alt locations (in relation with real word geometry). Built upon the well-established and extensively covered visual positioning system, GeoSpatialAPI offers mature functionality, covering nearly all areas included in the Google street view database.
\subsection{Placing the 3D virtual content}
Users have three options of architectural models to place and manipulate that are simple design, moderate design and complex design. Simple design has less number of polygons and components/objects (walls, roof, windows and doors) with no vegetation as compared to moderate and complex design. Moreover, textures used in simple design were of low resolution as compared to the textures used in moderate and complex design. Moderate design is more complex than the simple design and less complex than the complex design in similar manner. When user place a model on anchor, simple design is selected by default. However, for testing the other designs, users have to clear the screen and then load the second desired model, either moderate or complex one. After that, user can place it again by setting an anchor and repeat the procedure for testing other models. User interface for the implemented framework is shown in Figure \ref{fig:GeoPosePlacement}.
\subsection{Manipulating 3D virtual content}
Users can manipulate the architectural model by one-finger slide gesture for translating it along x, y and z axis. For scaling the models up and down, two-finger pinch gestures are used Moreover, twist gestures are used to rotate the models around x, y and z axis. Using these slide, pinch and rotate gestures, user can manipulate all the 7DOF (3 for translation, 3 for rotation anf 1 for scaling). %Figure \ref{fig:manipulation} shows the manipulation of 3D building models.
%\begin{figure}[t]  
%\centering
%\includegraphics[height=4.0in, width=6.5in]{images/manipulation.jpg}  
%\includegraphics[width=1\linewidth]{images/manipulation.jpg} \caption{(a). Scaling up 3D building model using two-finger pinch gesture. (b). Scaling down 3D building model using two-finger pinch gesture. (c). Translating 3D building model along x, y and z axis using one-finger slide gesture. (d). Rotating 3D building model around x, y, z axis using two-finger twist gesture.}
%\label{fig:manipulation}
%\end{figure}
   
\subsection{Additional features of Marker tracking}
\begin{figure}[h]  
\centering
\includegraphics[height=3in, width=2in]{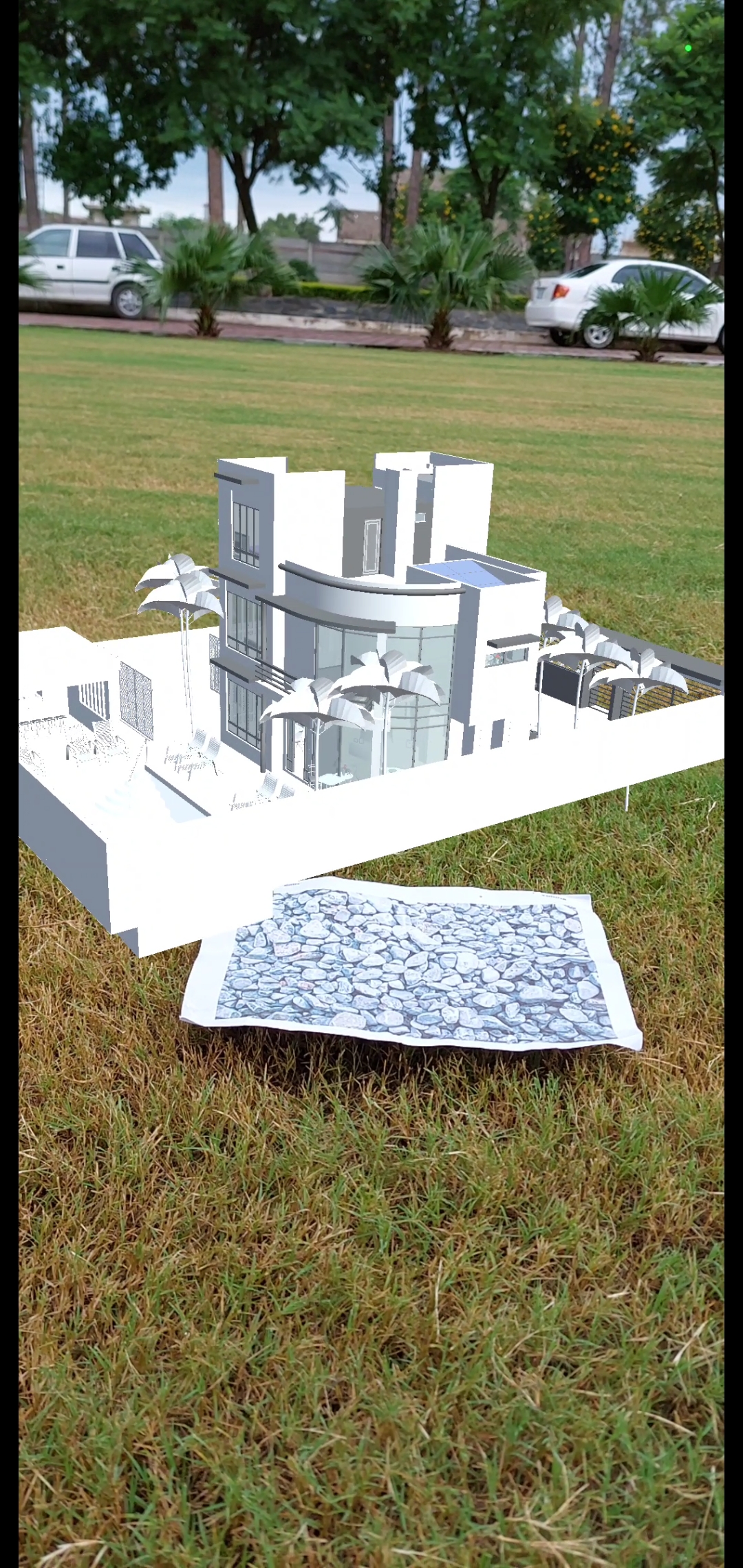}  
\caption{3D building model placed on a fiducial marker.}
\label{fig:markerPlacement}
\end{figure}

As an engineering feature, our framework is extendable with new features. For example, we are interested to compare our system with marker-based applications. For this purpose, the proposed system can be switched to/from marker-based HAR and vice versa. For marker-based HAR, we have used Vuforia Engine alongside ARCore, Unity3d and Microsoft Visual Studio to overlay architectural model onto some marker. User have to scan the fiducial marker provided to them and a building model is overlaid onto it. User can then manipulate it by scaling, rotating and translating it. User have to use one-finger slide gesture to translate the model along x, y and z axis, two-fingers pinch gesture to scale the model and two-finger twist gesture to rotate it around x, y and z axis just like markerless GHAR. The difference is that marker-based GHAR is confined to a marker for the placement of virtual content while merkerless GHAR is markerless. Figure \ref{fig:markerPlacement} shows an illustration of the technique.

\subsection{Markerless GHAR and Marker-based GHAR}  
The proposed system is named as GHAR with its key feature of markerless tracking. However, it also supports marker-based tracking. Therefore, we used markerless GHAR for marker-less HAR which provides markerless tracking and marker-based GHAR for marker-based HAR which provides marker-based tracking. According to our hypothesis the markerless GHAR is highly usable, easy to manipulate and understand compared to marker-based GHAR.

%%%%%%%%%%%%%%%%%%%%%%%%%%%%%%%%%%%%%%%%%%
\section{Evaluation paradigm for Proposed Framework}
\label{evaluation}
We assessed the proposed system through a subjective analysis employing two standard methods: the Handheld Augmented Reality Usability Study (HARUS) \cite{HARUS} and the System Usability Scale (SUS) \cite{SUS}. HARUS comprises 16 statements addressing manipulability and comprehensibility, while SUS consists of 10 statements evaluating usability. In general, three independent variables manipulability, comprehensibility, and usability were assessed, and the results were utilized for a subjective analysis of our proposed framework. Additionally, we measured numerical results, including errors and time. An independent sample statistical t-test \cite{Ttest} was employed to evaluate the reliability of the results. 
\subsection{Hypotheses}  To evaluate the three parameters, we formulated the following three null and corresponding three alternative hypotheses.
   \begin{enumerate}
    \item[A:] Null Hypotheses
   \begin{enumerate}
    \item[$H_01:$] There is no significant difference in terms of manipulability between markerless and marker-based GHAR. 
    \item[$H_02:$] There is no significant difference in terms of comprehensibility between markerless and marker-based GHAR.
    \item[$H_03:$] There is no significant difference in terms of usability between markerless and marker-based GHAR. 
\end{enumerate}  
 \item[B:] Alternate Hypotheses:
    \begin{itemize}
    \item[$H_a2:$] Markerless GHAR significantly outperforms marker-based GHAR in terms of manipulability. 
     \item[$H_a3:$] Markerless GHAR significantly outperforms marker-based GHAR in terms of comprehensibility.
      \item[$H_a4:$] Markerless GHAR significantly outperforms marker-based GHAR in terms of usability.
    \end{itemize} 
\end{enumerate}  
\subsection{Experimental Setup}
 We conducted a comparative study between markerless Handheld Augmented Reality (HAR)  and marker-based HAR through a user study. The experimental design employed the Randomized Posttest-Only Control Group Design \cite{researchDesign}. Two treatment groups, each comprising 20 participants (10 males, 10 females) aged between 22-32 years, were formed. One group experienced markerless GHAR, while the other experienced marker-based GHAR. Among the 40 participants, only 3 had prior experience with augmented reality applications in their daily work. Participants were introduced to the frameworks through a video demonstration, followed by practice sessions. After becoming familiar with the functionality, they were tasked with placement and manipulation (scaling, rotating, and translating) of 3D building artifacts using the provided frameworks. Following these activities, participants completed a questionnaire.  
 \begin{figure*}[!htbp]
% \vskip -5.5cm
    \centering  
        \includegraphics[width=0.74949\linewidth]{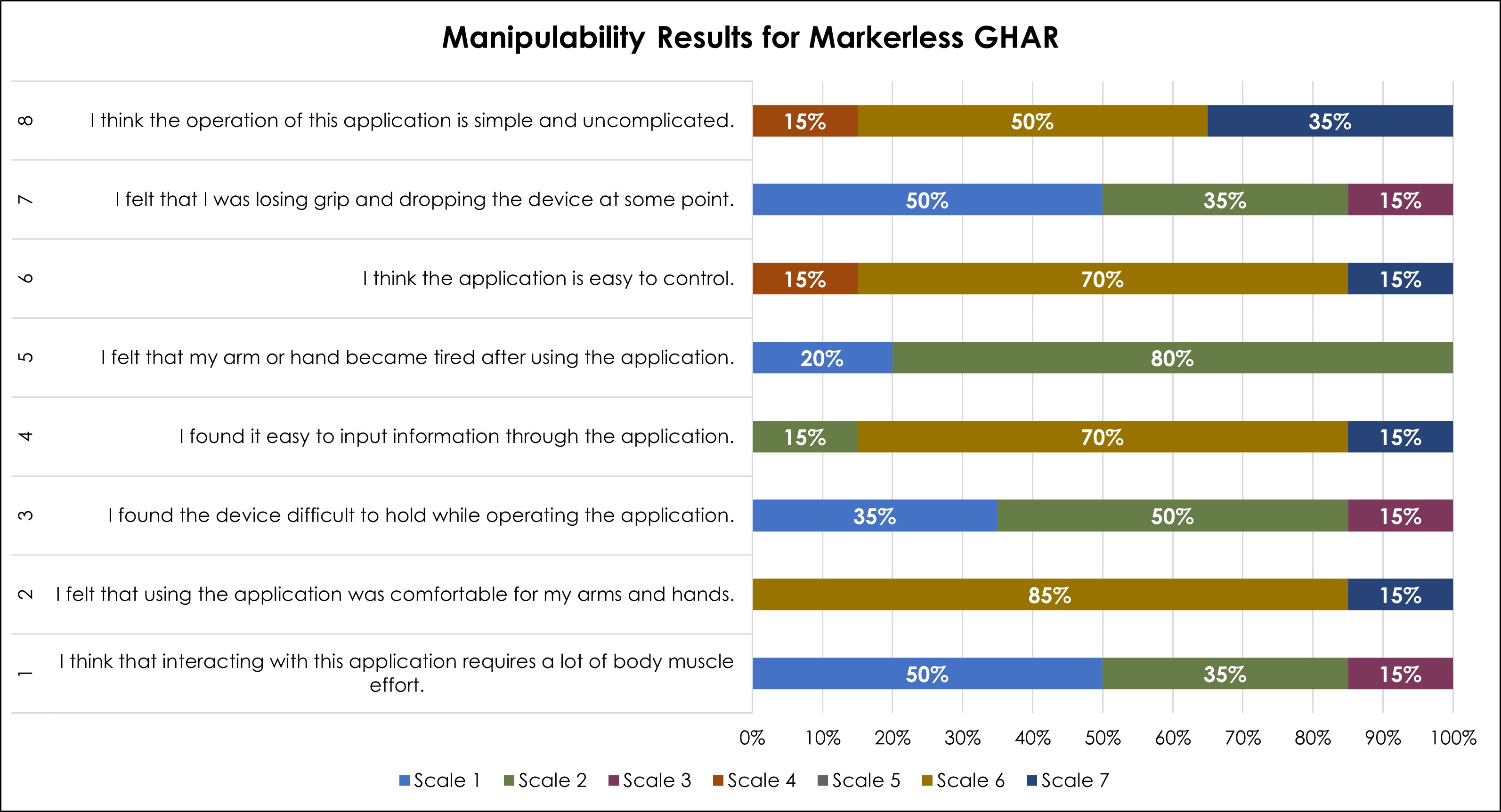}  
        \includegraphics[width=0.74949\linewidth]{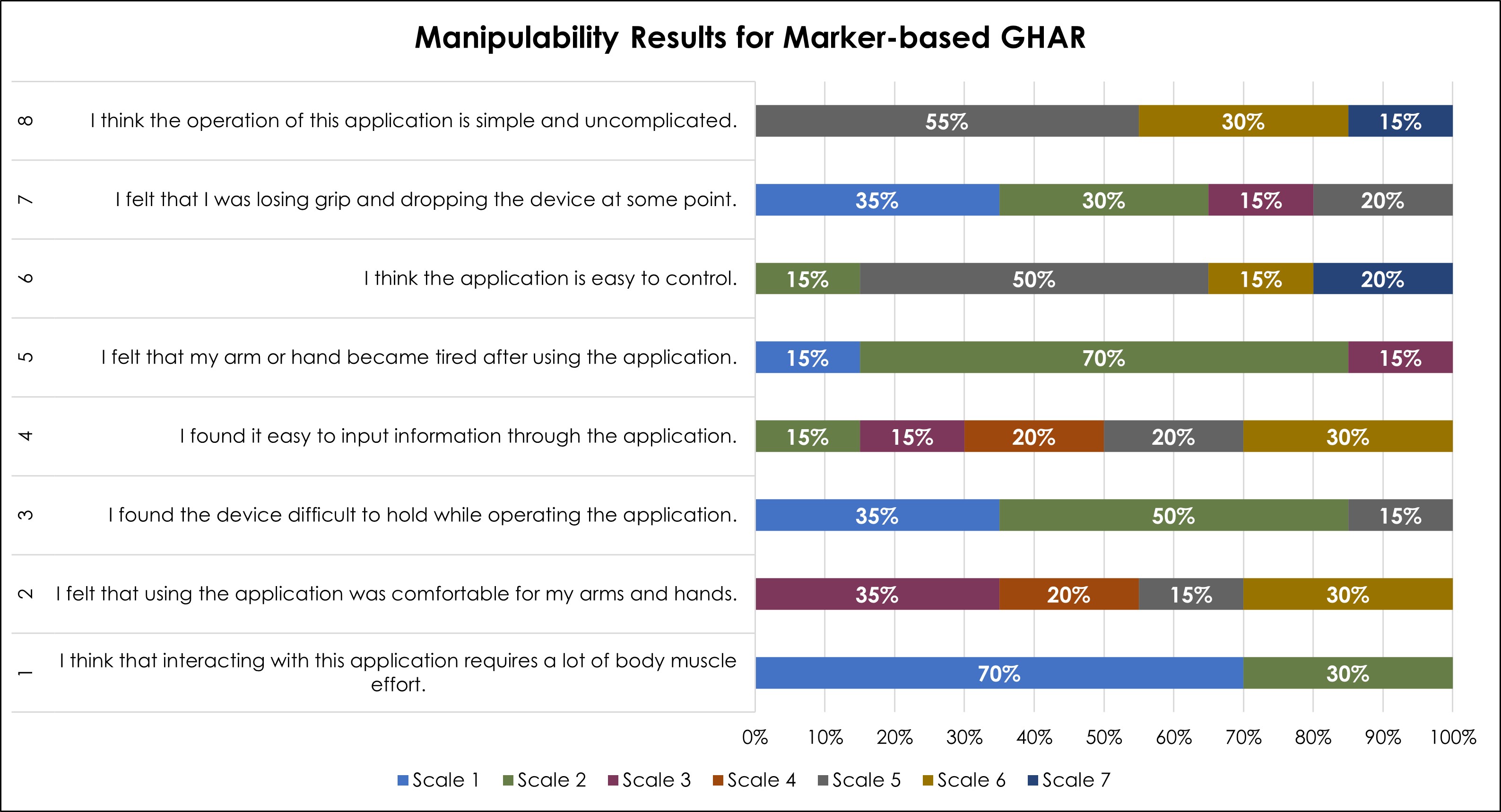}  
    \caption{HARUS Score distribution of manipulability measure for   Markerless GHAR (top) and   Marker-based GHAR (bott). x-axis represents the 8 statements regarding manipulability measure of HARUS while y-axis represents the average score distribution of individual statement on Scale 1-7 (where Scale 1 = Strongly Disagree and Scale 7 = Strongly Agree) in percentage.}
    \label{fig:Manipulability}
\end{figure*} 
 
\subsection{Sampling} 
In our experimental research, the individuals belonging to computing and civil infrastructure domains who use augmented reality supported handheld devices, comprised the population. Civil engineers, QA analysts, structural BIM specialists, MS Computer science research students and MS \& BS Software Engineering students are our accessible population. From this accessible population, we randomly selected a sample of 20 participants (10 females \& 10 males) for both the groups each. Among those 40 participants, we had civil engineers, QA analyst, structural BIM specialist, MS Software Engineering and BS Software Engineering student, and MS Computer Science research students. 

\subsection{Experimental Task}
Participants were given tasks to use either the marker-based or markerless GHAR framework. Each participant had to position a simple architectural design in suggested locations and adjust its placement until it reaches a required final position. They were also asked to visually explore the model by moving around it in 360 degrees. This process was repeated for two more architectural models with varying complexity (moderate and higher). After completing these tasks, participants from both groups were given the questionnaire to fill out. 
\begin{itemize}
    \item Task 1: Position a virtual 3D object in a particular location within your physical surroundings (markerless GHAR) or onto a marker, printed on A4 size paper (markerbased GHAR). 
    \item Task 2: Manipulate the 3D architectural model with multiple degrees of freedom (e.g., scale, rotate, translate) using pinch, twist, and slide gestures to align it precisely with a real-world structure.
    \item Task 3: Manipulate the 3D architectural model by isolating scaling, rotation and translation.  
    \item Task 4: Visually explore the model by moving around it in 360$^{\circ}$. Inspect its details and features.
    \item Task 5: Repeat tasks 1 to 4 for different types of  architectural models available (simple, moderate and complex design).
\end{itemize}

\section{Results}
\label{results}
On the basis of participants' responses, got from questionnaire survey, manipulability, comprehensibility and usability of the proposed framework is evaluated.

\subsection{Results for Manipulability}
This measure constitutes the 50\% of total HARUS score and comprised of 6 statements. HARUS score computed for the group treated with markerless GHAR is 42.40 while the group treated with marker-based GHAR gives a score of 37.14. Contribution of each statement in the score is illustrated in Figure \ref{fig:Manipulability}.  

\begin{figure*}[t]
\centering 
        \includegraphics[width=0.74949\linewidth]{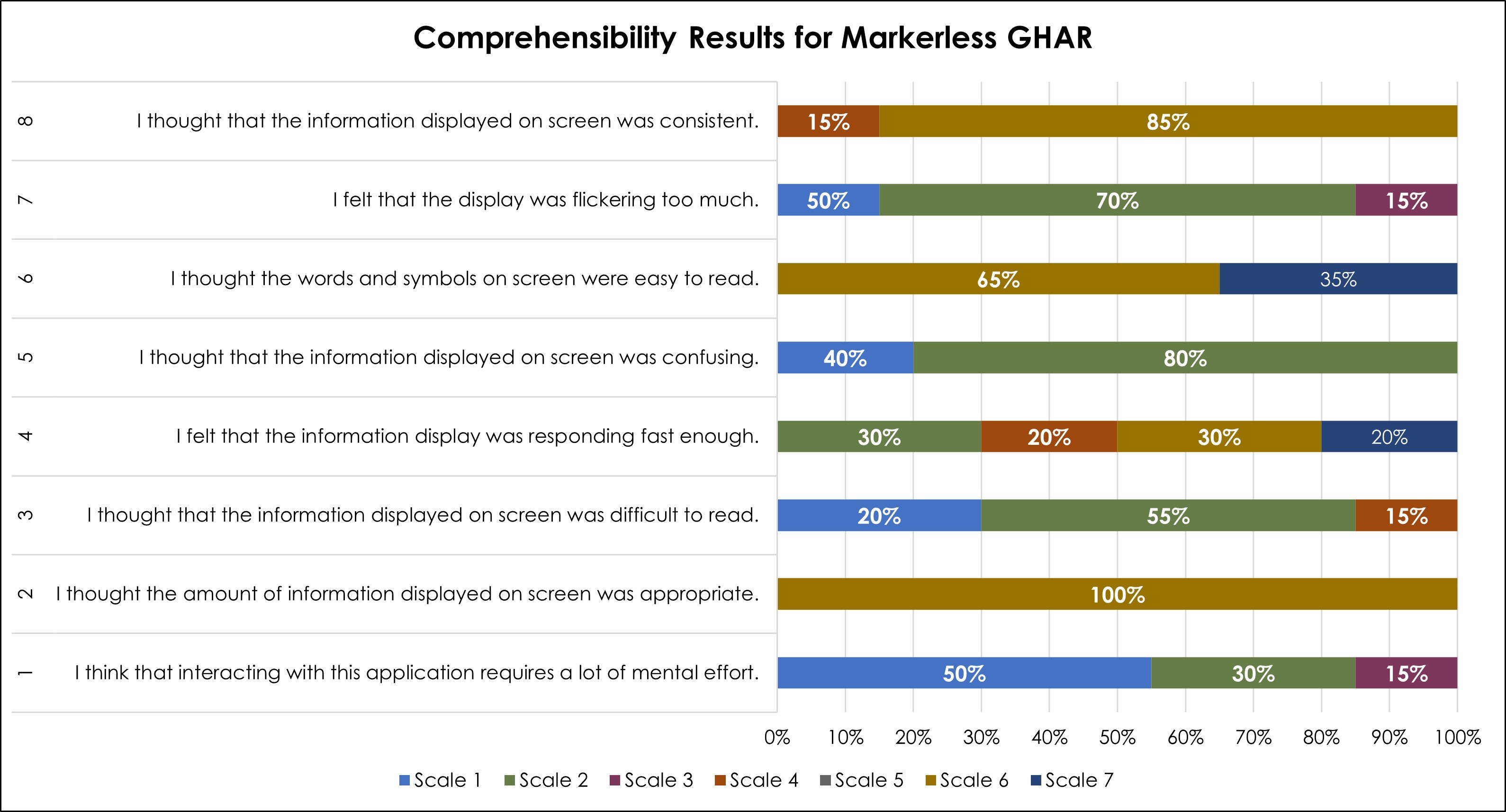}  \\
        \includegraphics[width=0.74949\linewidth]{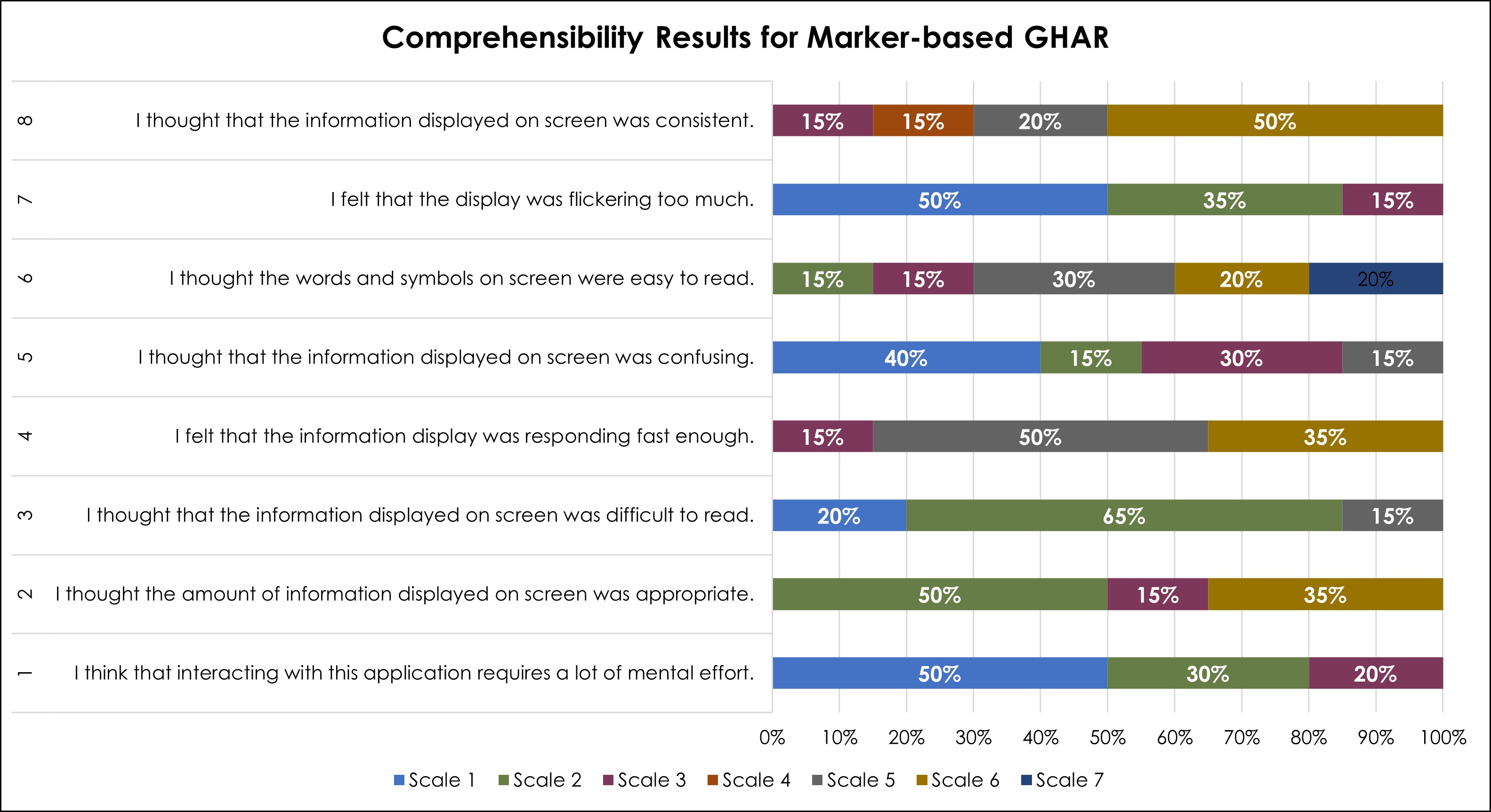} 
    \caption{HARUS Score distribution of comprehensibility measure for  Markerless GHAR and (top), and Marker-based GHAR (bottom). x-axis represents the 8 statements regarding comprehensibility measure of HARUS while y-axis represents the average score distribution of individual statement on Scale 1-7 (where Scale 1 = Strongly Disagree and Scale 7 = Strongly Agree) in percentage. }
    \label{fig:Comprehensibility}
\end{figure*} 
In summary, 24\% of markerless GHAR participants and 19\% of marker-based GHAR participants felt significant muscle effort during interaction. Comfort levels were high, with 88\% in markerless GHAR and 63\% in marker-based GHAR finding it comfortable for their arms and hands. Holding difficulty was reported by 26\% in markerless GHAR and 30\% in marker-based GHAR. Inputting information was considered easy by 79\% in markerless GHAR and 62\% in marker-based GHAR. Post-usage fatigue was noted by 26\% in markerless GHAR and 29\% in marker-based GHAR. Control ease was reported by 84\% in markerless GHAR and 73\% in marker-based GHAR. Losing grip and dropping the device were mentioned by 24\% in markerless GHAR and 34\% in marker-based GHAR. Operating the framework was considered simple by 86\% of markerless GHAR and 80\% of marker-based GHAR participants.  

\subsection{Results for Comprehensibility}
Comprehensibility constitutes the other half 50\% of HARUS score. Markerless GHAR treatment group has an HARUS score of 40.89 while the HARUS score of marker-based GHAR treatment group is 33.54. Contribution of each statement in the score is illustrated in Figure \ref{fig:Comprehensibility}. Here,   
 23\% of markerless GHAR participants and 24\% of marker-based GHAR participants found the framework requiring a higher mental effort. Information appropriateness was noted by 86\% in markerless GHAR and 51\% in marker-based GHAR. Reading difficulty was reported by 29\% in markerless GHAR and 32\% in marker-based GHAR. Fast responsiveness was acknowledged by 66\% in markerless GHAR and 72\% in marker-based GHAR. Information-induced confusion was experienced by 26\% in markerless GHAR and 34\% in marker-based GHAR. Symbols and words readability was affirmed by 91\% in markerless GHAR and 69\% in marker-based GHAR. Display flickering concern was raised by 29\% in markerless GHAR and 24\% in marker-based GHAR. Information consistency was agreed upon by 81\% in markerless GHAR and 72\% in marker-based GHAR.

\subsection{Results for Usability} 
\begin{figure*}[t] 
\centering 
        \includegraphics[width=0.7495\linewidth]{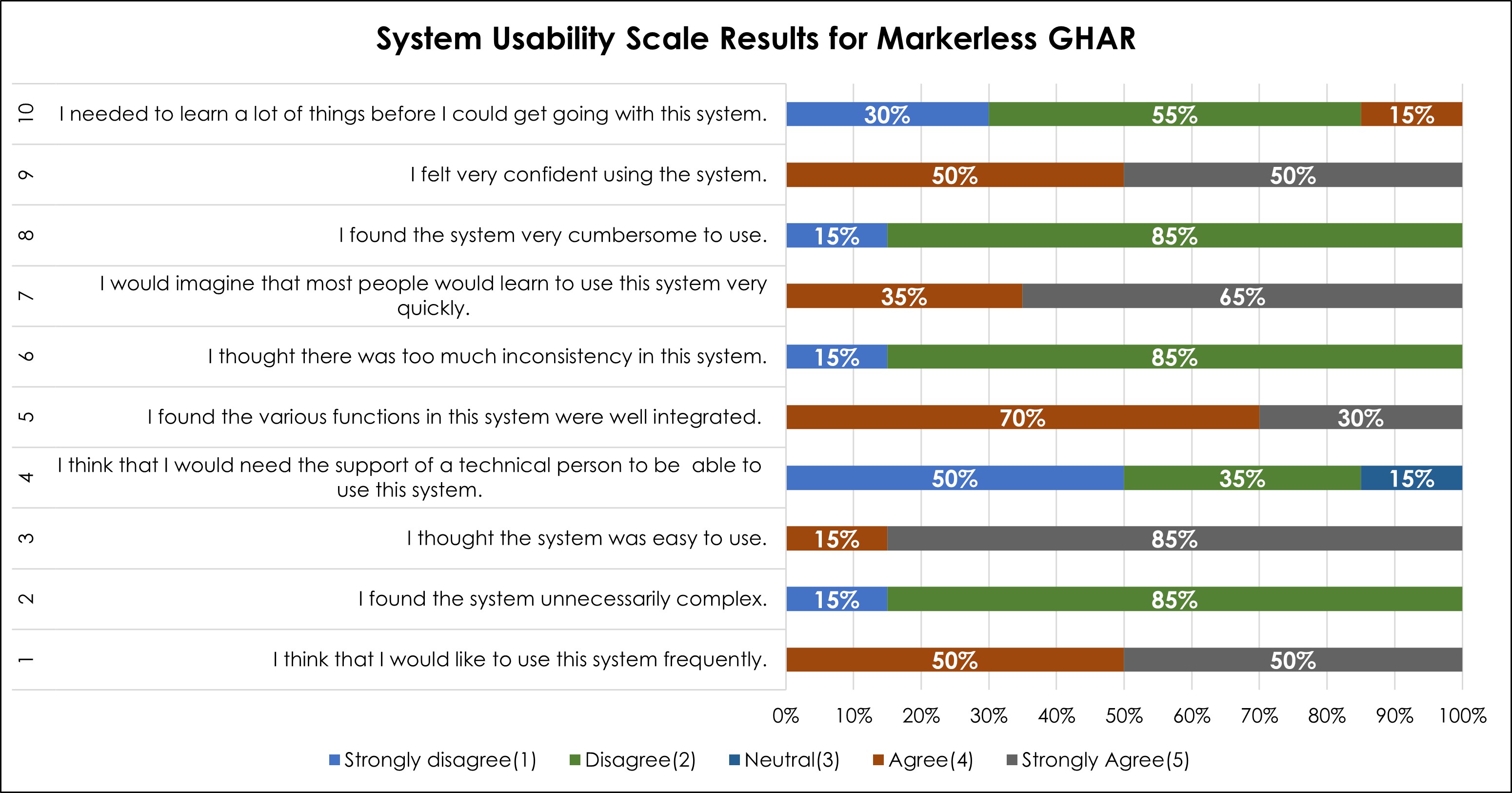} \\
        \includegraphics[width=0.7495\linewidth]{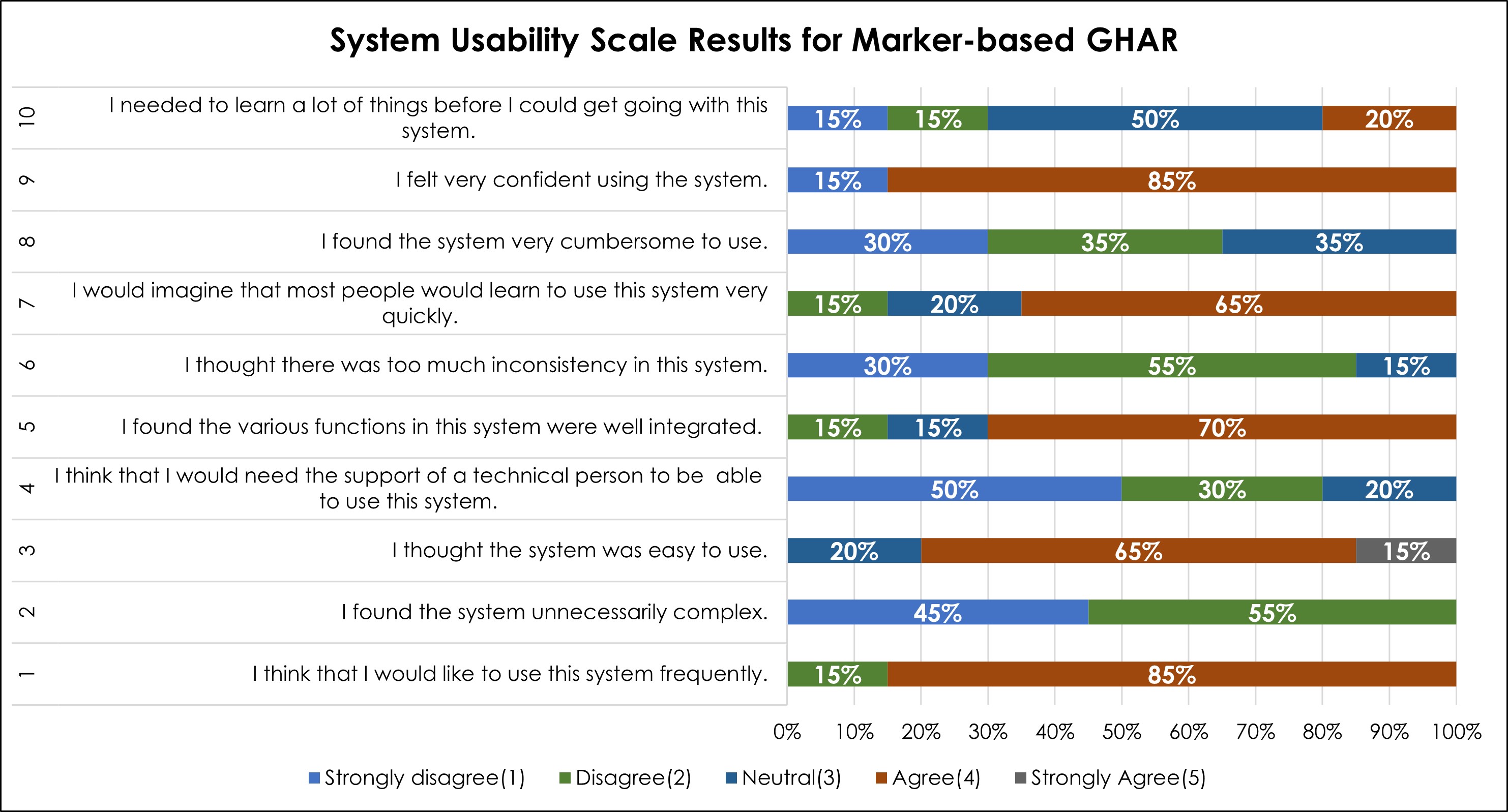} 
    \caption{SUS score distribution of usability measure for  Markerless GHAR and (Top), and  Marker-based GHAR (Bottoms). x-axis represents the 10 statements regarding usability measure of SUS while y-axis represents the average score distribution of individual statement on Scale 1-5 in percentage. }
    \label{fig:Usability}
\end{figure*}
SUS score for markerless GHAR treatment group is 84 and marker-based GHAR treatment group is 70.86. Contribution of each statement of SUS in its score is illustrated in Figure \ref{fig:Usability}. \\
For further details, 90\% participants of markerless GHAR treatment group and 74\% participants of marker-based GHAR treatment group agreed that they would like to use the framework frequently. 37\% participants of markerless GHAR treatment group and 31\% participants of marker-based GHAR treatment group found it complex. Framework was easy to use for 97\% participants of markerless GHAR treatment group and 79\% participants of marker-based GHAR treatment group. 33\% participants of markerless GHAR treatment group and 34\% participants of marker-based GHAR treatment group stated that they would need support of a technical person to be able to use the framework. According to 86\% participants of markerless GHAR treatment group and 71\% participants treated with marker-based GHAR, various functions of the framework were well integrated. 37\% participants of markerless GHAR treatment group and 37\% participants of marker-based GHAR treatment group said that they found inconsistency in the framework. 93\% markerless GHAR treatment group participants and 70\% marker-based GHAR treatment group participants agreed that most of the people would learn to use this system very quickly. 37\% participants of markerless GHAR treatment group and 41\% participants of marker-based GHAR treatment group found the framework cumbersome to use. 90\% participants of the group treated with markerless GHAR and 71\% participants treated with marker-based GHAR said that they felt confident while using the framework. 40\% participants of markerless GHAR treatment group and 55\% participants of marker-based GHAR treatment group stated that they needed to learn a lot of things before they could get going with the system.

\subsection{Overall Result Analysis}
Table \ref{tab:HARUS&SUS} shows that HARUS score for markerless GHAR is 82.76 and for marker-based GHAR, HARUS score is 72.34. It is evident from the results that markerless GHAR outperforms marker-based GHAR. Similarly, SUS score for markerless GHAR that is 84 depicts that it outperforms marker-based GHAR with SUS score of 70.86 in terms of usability.

% Table generated by Excel2LaTeX from sheet 'Sheet2'
\begin{table}[htbp]
  \centering
  \caption{Summary of results. Comparing markerless vs marker-based GHAR. }
    \resizebox{\linewidth}{!}{
    \begin{tabular}{|l|c|c|c|c|}
    \hline
   % & \multicolumn{3}{|c|}{Handheld Augmented Reality Usability Scale } & System Usability Scale (SUS) \\
    & \multicolumn{3}{|c|}{HARUS} &SUS  \\
    \hline
          & \multicolumn{1}{p{7.0em}|}{Manipulability Score} & \multicolumn{1}{p{8.5em}|}{Comprehensibility Score} & \multicolumn{1}{p{3.8em}|}{HARUS Score} & SUS Score \\
    \hline
    Markerless GHAR & 42.40 & 40.89 & 86.76 & 84 \\
    %\textbf{} &  &  &  &  \\
    \hline
    Marker-based GHAR & 37.14 & 33.54 & 73.63 & 70.86 \\
    %\textbf{} &  &  &  &  \\
    \hline
    \end{tabular}%
  \label{tab:HARUS&SUS}%
}
\end{table}%

\subsection{Statistical T-test Results}
As evident from the Table \ref{tab:ttest}, we can reject the null hypotheses and can say that the difference in the means of both groups does not happen by chance. 
\begin{itemize}
    \item Markerless GHAR (Geopose-based tracking) significantly better manipulability than marker-based GHAR t(38) = 3.04, p = 0.004.
    \item Markerless GHAR (Geopose-based tracking) has significantly better comprehensibility than marker-based GHAR t(38) = 4.89, p = 0.0.
    \item Markerless GHAR (Geopose-based tracking) also has significantly better usability than marker-based GHAR t(38) = 3.82, p = 0.0.
\end{itemize}   
\begin{table}[h]
  \centering
  \caption{Results for Independent Sample T-test.}
    \begin{tabular}{|l|c|c|c|c|}
		\hline
        Variables   &  t  &   df &    Sig.(2-tailed)  &  Cohen's d  \\ 
        \hline
    Usability  & 3.82  & 38  & 0.000  & 1.2 \\ \hline
     Manipulability  & 3.04  & 38  & 0.004  & 1.0 \\ \hline 
     Comprehensibility  & 4.89  & 38  & 0.000  & 1.5 \\
      % &   &   &  &  \\
     \hline
    \end{tabular}%
  \label{tab:ttest}%
\end{table}% 

\section{Discussion}
\label{discussion}
The comparative analysis of markerless and marker-based GHAR systems reveals significant differences in user experience, highlighting the advantages of markerless GHAR as a more user-friendly and effective solution for handheld AR applications. The HARUS scores, which evaluate physical interaction and cognitive comprehensibility, favor the markerless system, suggesting its superior design and usability.

In terms of physical interaction and comfort, the markerless GHAR system provides a more ergonomic experience. Users reported lower muscle effort and higher comfort levels in their arms and hands compared to the marker-based system. This indicates that the markerless system is less physically demanding, making it suitable for extended use. Additionally, participants found inputting information easier with the markerless system, reflecting its intuitive design. Control ease and reduced concerns about dropping the device further emphasized the advantages of the markerless approach, which fosters confidence and minimizes operational challenges.

Cognitive comprehensibility also favored the markerless system. Users found the information presented by the markerless framework to be more appropriate and relevant, contributing to a more seamless and intuitive experience. Although responsiveness was slightly better in the marker-based system, the markerless system excelled in reducing confusion and improving the readability of symbols and text. These findings suggest that the markerless system is better equipped to meet the cognitive demands of users, facilitating a smoother interaction with the AR interface.

The usability and adoption metrics strongly support the markerless system as the preferred choice. Participants expressed a higher willingness to use the markerless system frequently and found it easier to operate. Users also felt more confident while using the markerless system and appreciated the integration of its various functions. In conclusion, the markerless GHAR system offers a superior user experience compared to the marker-based system, particularly in terms of comfort, ease of use, and comprehensibility.

\section{Conclusion}
\label{conclusion}
 In this research work, we have presented the handheld augmented reality framework \textit{GHAR} in two modes: Markerless GHAR and Marker-based GHAR, designed for the convenient positioning and manipulation of 3D virtual objects. We utilized 3D architectural design models of civil infrastructure as a use case to implement our framework. For Markerless GHAR, GeoPose has been employed to position the 3D virtual objects with respect to the globe, considering both the position and orientation of the device. This approach allows users to focus more on the task without being concerned about markers while manipulating different degrees of freedom (DOFs) of virtual objects. We conducted an evaluation of the usability, manipulability, and comprehensibility of Markerless GHAR against Marker-based GHAR, which utilizes a fiducial marker to overlay 3D virtual content. Similar to Markerless GHAR, this mode also employs a 3D architectural model of civil infrastructure that can be overlaid on a detected marker, requiring users to manipulate seven DOFs to control the model. Results show that Markerless GHAR significantly outperforms Marker-based GHAR in terms of usability, manipulability, and comprehensibility.\\
\textbf{Future Work:} In the future, we aim to implement our framework for head-mounted displays to assess the impact of the field of view on results. Additionally, we plan to test this framework in other large-scale applications to determine its optimal fields of application. Exploring its effects in different locations, altitudes, and headings for location-based applications is also on our agenda. Given its generic nature, the framework offers ample opportunities for extension and utilization in various use cases and application areas.

%\IEEEraisesectionheading{\section{Introduction}}
% \input{content/01-introduction}
% \input{content/02-Relatedwork}
% \input{content/03-Method}
% \input{content/04a-ExpSetup}
% \input{content/04b-Results} 
% % \input{content/05-Demo}
% \input{content/06-Discussion} 

% \section*{Acknowledgments}
% This should be a simple paragraph before the References to thank those individuals and institutions who have supported your work on this article.


% \section{References Section}
% You can use a bibliography generated by BibTeX as a .bbl file.
%  BibTeX documentation can be easily obtained at:
%  http://mirror.ctan.org/biblio/bibtex/contrib/doc/
%  The IEEEtran BibTeX style support page is:
%  http://www.michaelshell.org/tex/ieeetran/bibtex/
 
% You can manually copy in the resultant .bbl file and set second argument of $\backslash${\tt{begin}} to the number of references
%  (used to reserve space for the reference number labels box).

%\begin{thebibliography}{1}
\bibliographystyle{IEEEtran}  
\bibliography{refs}
%\end{thebibliography}

% \newpage

 

% \end{IEEEbiography}


%\vfill
% \vspace{-33pt} 
% \input{biography.tex}

\vfill
\end{document}